\documentclass[reprint,aps,prmaterials,longbibliography,superscriptaddress]{revtex4-2}
\usepackage{amsmath,amssymb}
\usepackage{graphicx}
\usepackage[dvipsnames]{xcolor}
\usepackage{dsfont}
\usepackage[T1]{fontenc}

\newcommand*{\rom}[1]{\expandafter\@slowromancap\romannumeral #1@}

\begin{document}
\title{Signatures of bulk g-wave altermagnetism in optically excited $\alpha$-MnTe}

\author{Luca~Felipe~Haag}
        \affiliation{Department of Physics and State Research Center OPTIMAS, RPTU University Kaiserslautern-Landau, 67663 Kaiserslautern, Germany}
\author{Marius~Weber}
        \affiliation{Institute of Physics, Johannes Gutenberg University Mainz, 55099 Mainz, Germany}
        \affiliation{Department of Physics and State Research Center OPTIMAS, RPTU University Kaiserslautern-Landau, 67663 Kaiserslautern, Germany}
\author{Kai~Leckron}
    \affiliation{Department of Physics and State Research Center OPTIMAS, RPTU University Kaiserslautern-Landau, 67663 Kaiserslautern, Germany}
\author{Libor~\v{S}mejkal}
    \affiliation{Max Planck Institute for the Physics of Complex Systems, 01187 Dresden, Germany}
    \affiliation{Max Planck Institute for Chemical Physics of Solids,  01187 Dresden, Germany}
    \affiliation{Institute of Physics, Academy of Sciences of the Czech Republic, Cukrovarnick\'{a} 10, Prague 6, Czech Republic}
\author{Jairo~Sinova}
    \affiliation{Institute of Physics, Johannes Gutenberg University Mainz, 55099 Mainz, Germany}
    \affiliation{Department of Physics, Texas A\&M University, College Station, Texas 77843, USA}
\author{Hans~Christian~Schneider}
    \affiliation{Department of Physics and State Research Center OPTIMAS, RPTU University Kaiserslautern-Landau, 67663 Kaiserslautern, Germany}
    
\date{\today}
\begin{abstract}

For planar d-wave altermagnets, it has been shown that a spin polarization can be induced in a controlled fashion by ultrashort-pulse excitation, even though the material is magnetically compensated. Here, we theoretically analyze the response of the prototypical bulk $g$-wave altermagnet $\alpha$-MnTe to linearly polarized ultrashort pulses. We demonstrate how the electronic spin response in $\alpha$-MnTe exhibits different symmetry characteristics by calculating the excited electron distributions based on \emph{ab initio} band structure data. These characteristics depend not only on the nodal planes of the bulk g-wave altermagnet, but also on the excitation pulse. We present a simple procedure to analyze the excited-state characteristics via two-dimensional cuts through the three-dimensional Brillouin zone, which can be used as guidance to present-day magneto-optical techniques.
\end{abstract}

\pacs{}

\maketitle

\section{Introduction \label{ch:section1}}

Altermagnets promise new magnetic functionalities due to their spin-dependent electronic properties favorable for spintronics applications. These include not only the combination of ferromagnet-like splitting in k-space, antiferromagnet-like zero net magnetization and THz dynamics, but also unique responses to electric and magnetic fields due to their nodal d, g, or i-wave spin order~\cite{altermagnet1,altermagnet2}. This is because their spin-split bands in momentum space make it possible to efficiently control electronic spins while their vanishing net magnetic moment protects the spin information against unwanted external disturbances. A wide variety of altermagnetic candidate materials, including insulators and semiconductors, but also metals, is currently being screened for their use in spintronics~\cite{altermagnet2,reichlova_nature_review,mokrousov_review_altermagnets,guo2025}. 

The semiconducting compound~$\alpha$-MnTe~\cite{altermagnet1} has emerged as a prominent material for the study of altermagnetism as its nodal bulk g-wave spin order has been directly observed in the spin-dependent band splitting by angle-resolved photo emission spectroscopy (ARPES)~\cite{krempasky2024,osumi_arpes_mnte2024,Arpes_japan1}. This is in contrast to another popular altermagnetic candidate material RuO$_2$, for which differing results have been published for different structures~\cite{fedchenko_ruo2,AHE_ruO2_china1,moser_ruo2,muSR_ruO2_japan}. In the case of $\alpha$-MnTe, its altermagnetic character has also been experimentally confirmed through anomalous Hall effect measurements~\cite{betancourt_MnTe} and anisotropic magneto resistance measurements \cite{GonzalezBetancourt2024_anisotropic_magres}. Further, X-ray magnetic circular dichroism has been used to characterize its magnetic properties~\cite{amin_xmcd_mnte_2024}, and a series of theoretical and experimental investigations have focused on its transport characteristics~\cite{kazmin2025andreev,GonzalezBetancourt2024_anisotropic_magres,kluczyk_anomalous_hall}.

$\alpha$-MnTe is ideally suited to study the optical response on ultrafast timescales since samples with a domain size on the order of several microns have been fabricated~\cite{Amin2024_nature}. We recently showed using \emph{ab initio} based calculations for the planar d-wave altermagnets RuO$_2$ and KRu$_4$O$_8$ that non-equilibrium spin densities can be injected by \emph{linearly} polarized laser pulses~\cite{weber_optics_ruo2,weber_dynamics_kru4o8_newton}. Such a non-equilibrium effect can be detected using magneto-optical techniques even for N\'{e}el vector orientations for which static MOKE is forbidden by symmetry~\cite{mazin_2023_dielectric_function}. These theoretical predictions were validated by time-resolved magneto-optical Kerr measurements~\cite{weber_optics_ruo2} on RuO$_2$ thin films. The possibility to control the induced spin polarization in such a compensated material by the linear polarization state of the exciting laser pulse is a robust effect rooted in the altermagnetic symmetries that has also been obtained in other theoretical studies~ \cite{vila_orbital_prb,zhou2025polarizing,eskandari2025_prb,zhou2026_prb,li2026_nano}. In addition, phonon and magnon contributions, which are secondary effects following the optical excitation, have been identified on picosecond timescales in magneto-optical Kerr effect measurements depending on the light polarization~\cite{gray_trMOKE}. 

\begin{figure*}[t]
    \centering
    \includegraphics[width=\linewidth]{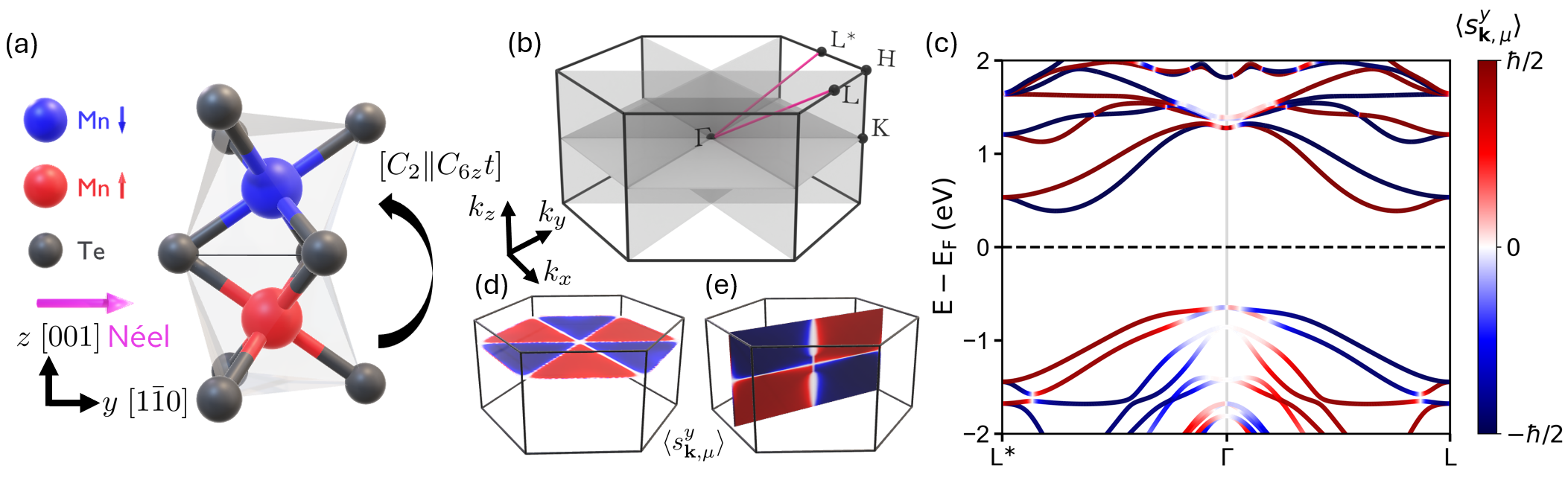}
    \caption{(a) Crystal structure of $\alpha$-MnTe together with the octahedral environment that breaks the pure translational and inversion symmetry between the magnetic sublattices. (b) Corresponding BZ in reciprocal space with high symmetry points and highlighted spin degenerate nodal plane structure (gray shaded planes). (c) 1D band structure along the pink high symmetry path visualized in b) where red and blue indicate spin-up and spin-down states, respectively, white signifies the spin-mixed regions due to SOC. (d) and (e) show spin cuts for different two-dimensional cross sections of the BZ, i.e., the computed spin expectation values $\langle s_{\boldsymbol{k},\mu}^{y}\rangle$ of the uppermost valence band color coded with the same colormap as (c).}
    \label{fig:fig1}
\end{figure*}

In this paper, we explore the spin dependent electronic response of a prototypical bulk g-wave altermagnet as it can be induced by ultrashort optical pulses, using $\alpha$-MnTe as a representative. Due to the additional vertical and horizontal nodal planes in a bulk g-wave structure compared to the planar d-wave case, one can expect the optically induced effects to differ in these two altermagnetic systems. We present a detailed study of the dependence of the spin polarization on polarization angle and wavelength of ultrashort pump pulses. The existence of high-quality samples should make it possible to explore these computed signatures of g-wave altermagnetism in future optical pump-probe experiments.

The paper is organized as follows. In Sec.~II we review the electronic ground-state properties of $\alpha$-MnTe, as they are obtained from density-functional theory (DFT) and form the basis for the subsequent calculations. In Sec.~III we investigate the excited-state spin polarization that can be induced by ultrashort optical pulses. We show how the symmetry of the spin polarization in $k$-space is dictated by the incidence angle of the pulse with respect to $[001]$-direction of the crystal structure. We then discuss the characteristics of the spin-dependent electronic response of this system. In Sec.~IV we analyze the photon energy dependence of the signatures found in Sec.~III. In Sec.~V we present our conclusion, and  give some details on the calculated dielectric functions in the Appendix~\ref{ch:dielectric}. Finally, we present results obtained by neglecting spin-orbit coupling (SOC) in the Appendix~\ref{ch:appendix2}. 

\section{Electronic Ground State \label{ch:section2}}

In this section, we present details of the electronic properties of $\alpha$-MnTe, which we will apply in the next section to determine the optical response. The single-particle band structure and Bloch wave functions were determined \emph{ab initio} by DFT with the ELK all-electron LAPW code~\cite{elk-code} using the PBE generalized gradient approximation (GGA)~\cite{PBE}. In order to account for the strongly correlated electrons of Mn, Hubbard parameters for the Mn $3d$-orbitals of $U=4.0$ eV and $J=0.97$ eV \cite{betancourt_MnTe} were applied within the fully localized limit (FLL) scheme of the ELK Code \cite{elk_fll_dftu, elk-code}. To ensure tight convergence, the calculations are carried out on a $26\times26\times16$ $k$-grid.
Figure~\ref{fig:fig1}(a) depicts the real-space structure of $\alpha$-MnTe, which crystallizes in the hexagonal space group P6$_3$/mmc \cite{kriegner_MnTe,betancourt_MnTe}. The lattice constants are $a=4.134\,\text{\AA}$ and $c=6.652$\,\AA. The magnetic character of the manganese (Mn) atoms is shown as red (spin up) and blue (spin down) with respect to the N\'{e}el vector (pink arrow) orientation. Importantly, the nonmagnetic tellurium (Te) atoms (gray) form an octahedral crystal environment that breaks the pure translational and inversion symmetry between the two magnetic sublattices.
For the DFT calculation, we set the N\'{e}el vector direction to the $[1\bar{1}0]$-direction of the crystal as shown Fig.~\ref{fig:fig1}(a), which is one of the easy axis orientations~\cite{kriegner_MnTe}. The converged structure exhibits a magnetic moment of $\pm4.447$ $\mu_{\mathrm{B}}$ per Mn atom in the unit cell with a slight canting of the Mn magnetic moments due to SOC resulting in a small ferromagnetic moment of $3.7\cdot10^{-4}$ $\mu_{\mathrm{B}}$ in the $[001]$-direction. These results align with those reported in previous studies~\cite{betancourt_MnTe,krempasky2024,mazin2024_MnTe}.
\begin{figure*}[t]
    \centering
    \includegraphics[width=\linewidth]{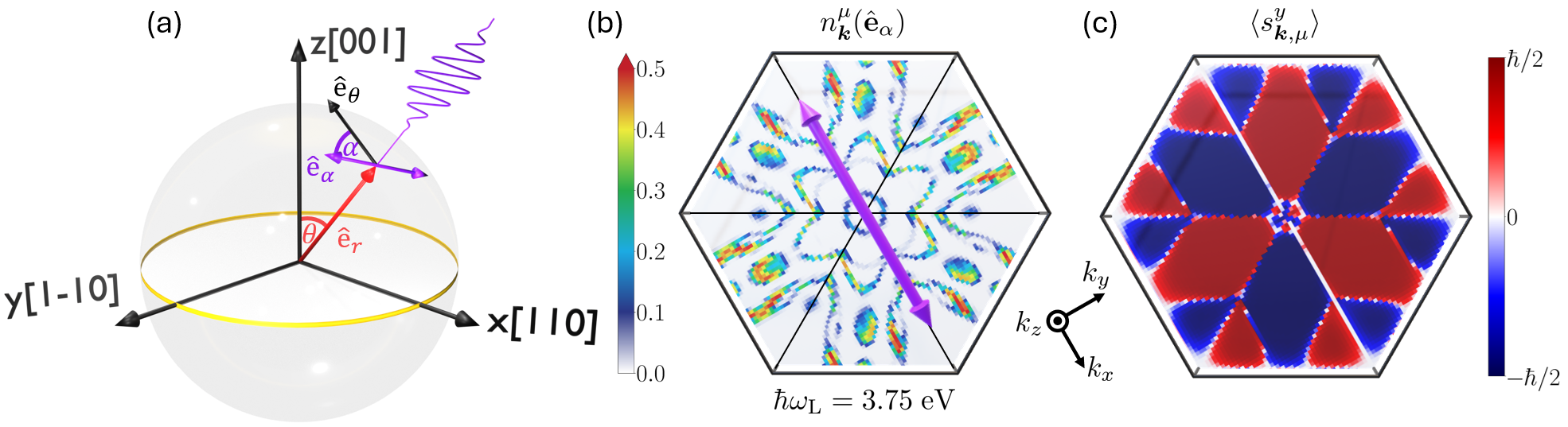}
    \caption{(a) Polarization direction $\hat{\mathbf{e}}_{\alpha}$ of the linearly polarized excitation E-field  and spherical unit vectors. (b) 2D cross section $k_z=0.5(\pi/c)$ of the excited electronic distribution in the first conduction band for the E-field vector of the laser pulse oriented along the $[001]$-direction of the crystal, i.e., $\theta=0^{\circ}$, with photon energy $\hbar\omega_{\mathrm{L}}=3.75$~eV for an excitation angle of $\alpha=0^{\circ}$. (c) Corresponding spin structure of the band.}
    \label{fig:fig2}
\end{figure*}

The spin-group~\cite{litvin, altermagnet1} used to describe the real-space and spin-space structure the $\alpha$-MnTe crystals is $^{2}6/^{2}m^{2}m^{1}m$. Its sublattices are connected by a translation along the $z$-axis by half a lattice vector combined with a $60^{\circ}$ rotation (screw symmetry) in real space accompanied by an inversion in spin space, which is conventionally denoted by $[C_2\Vert C_{6z}t_{1/2}]$. These symmetries give rise to four spin-degenerate nodal planes in $k$-space as illustrated in Fig.~\ref{fig:fig1}(b), which shows the Brillouin zone (BZ) of $\alpha$-MnTe. The three vertical nodal planes and one horizontal nodal plane (gray) constitute the defining feature of the bulk g-wave character. In addition to the nodal planes, we have represented in Fig.~\ref{fig:fig1}(b)  by pink lines the high symmetry path $L^{*}-\Gamma-L$, along which the band structure is plotted in Fig.~\ref{fig:fig1}(c). The color code of the bands indicates the spin expectation value $\langle s_{\boldsymbol{k},\mu}^{y}\rangle$ of the electronic Bloch state with crystal momentum~$\boldsymbol{k}$ and band index~$\mu$. The spin structure is described by the projection $s^y$ of the spin vector on the Néel vector direction~$+y$ with dark red indicating an alignment parallel and dark blue an orientation antiparallel to the $y$-axis. We will refer to these as ``spin-up'' and ``spin-down'' in the following. The white regions represent spin-mixed regions that arise due to SOC. It is evident from Fig.~\ref{fig:fig1}(c) that the bands along the paths $L^{\ast}-\Gamma$ and $\Gamma-L$ have equal dispersion but exhibit a telltale alternating spin pattern. We note that the spin degeneracy within the nodal planes is lifted due to the presence of SOC~\cite{krempasky2024}, but for simplicity we will refer to these as nodal planes in the following. More details about the lifting of the Kramers degeneracy can be found in Supplementary Information S3~\cite{supplement}.

Figures~\ref{fig:fig1}(d) and (e) show the spin expectation value $\langle s_{\boldsymbol{k},\mu}^{y}\rangle$ of the uppermost valence band for different two-dimensional cross sections of the BZ. We will call these plots ``spin cuts'' and use them as simplified pictures to augment our computed results below. As demonstrated for RuO$_2$~\cite{weber_optics_ruo2}, and will be checked for MnTe in the following, these static spin cuts may provide a simple and intuitive visualization of the optically excited spin polarization in the full BZ. In particular, Fig.~\ref{fig:fig1}(d) shows this spin expectation value as a function of $(k_x,k_y)$ in a plane located at $k_z= 0.5 (\pi/c)$. Figure~\ref{fig:fig1}(e) depicts the spin expectation value as function of $(k_y,k_z)$ in the plane at $k_x=0$. We have chosen the spin cuts for the topmost valence band because these bands do not exhibit crossings and therefore make the altermagnetic symmetry in the band structure well visible. The spin cut in Fig.~\ref{fig:fig1}(d) is partitioned by the three vertical nodal planes into six areas with alternating spin orientation. The regions opposite to each other, i.e., related by the mapping $(k_x,k_y,k_z)$ to $(-k_x,-k_y,k_z)$, show a reversed spin character. The spin cut of the $k_y$-$k_z$ plane in Fig.~\ref{fig:fig1}(e) is divided into four areas with alternating spin orientation where the spin character of areas opposite in $k$-space is now identical. This specific pattern resembles that of the spin cuts of planar d-wave altermagnets such as RuO$_2$ and KRu$_4$O$_8$ as shown in Ref.~\cite{altermagnet1}, for which we have demonstrated that the excited spin polarization can be controlled by the polarization of the linearly polarized laser pulse \cite{weber_optics_ruo2}. The spin cuts shown in~Figs.~\ref{fig:fig1}(d) and (e) suggest that such an effect should also arise for bulk g-wave altermagnets when non-equilibrium carrier distributions are excited by optical pulses. We will therefore study the consequences of an optical excitation in dependence on polarization and incident angle of the pulses in $\alpha$-MnTe for a fixed laser wavelength next.

\section{Optical Excitation Conditions: Angular Dependence \label{ch:section3}}

In planar altermagnets the qualitative dependence on the laser excitation can be well explained by considering a 2D cross section of the BZ \cite{weber_dynamics_kru4o8_newton}. To extend this analysis of the optical excitation to \emph{bulk} altermagnets with a more complex nodal-plane structure it is, however, crucial to investigate the optical excitation in the entire 3D BZ. Thus, all $k$-integrated signals in the following are evaluated on the full three-dimensional $26\times26\times16$ DFT $k$-grid.  

\begin{figure*}[t]
    \centering
    \includegraphics[width=\linewidth]{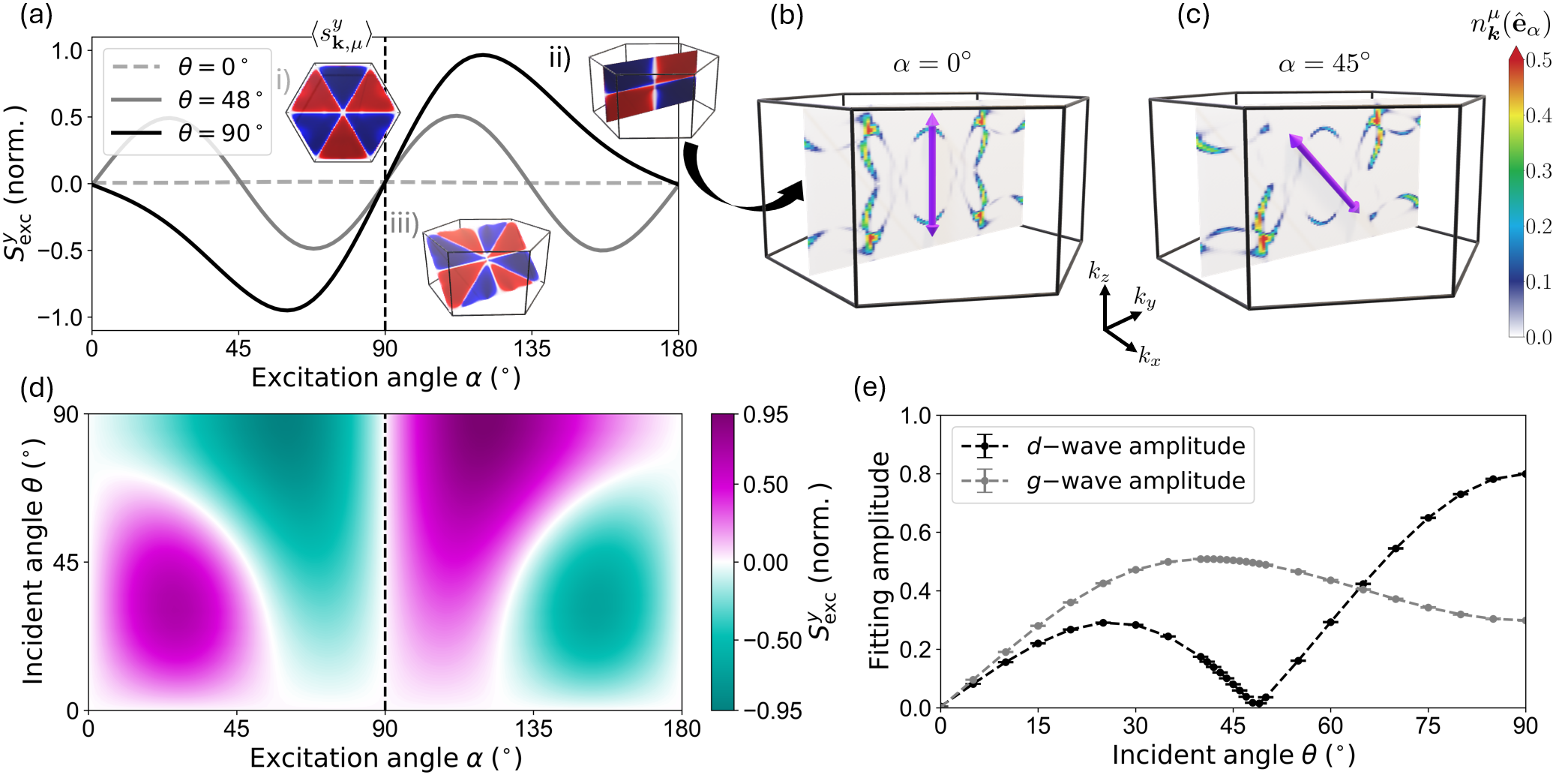}
    \caption{(a) Excited spin polarization as a function of the excitation angle~$\alpha$ for three different incident directions. The insets show the spin cuts of the uppermost valence band for a BZ cross-section perpendicular to the three incident directions, (i) $\theta =0^\circ$, (ii) $\theta = 48^\circ$ and (iii) $90^\circ$, respectively. (b), (c) Excited carrier distribution in the first conduction band for an incidence angle of $\theta=90^{\circ}$ in the corresponding perpendicular $k$-slice. Highlighted are the $\alpha=0^{\circ}$ (b) and $\alpha=45^{\circ}$ (c) polarization directions of the pulse, which correspond to a vanishing and a negative excited spin polarization, respectively. (d) Excited spin polarization as a function of incident angle $\theta$ and excitation angle $\alpha$. (e) Amplitudes of the sinusoidal fit function with d- and g-wave character for different incident directions as defined in Eq.~\eqref{eq:fit_function}. The error bars indicate the standard deviation of the fitting procedure (see Supplementary Information~S2).}
    \label{fig:fig3}
\end{figure*}

Our theoretical approach is as follows. We use the \emph{ab initio} electronic dispersion~$\epsilon_{\boldsymbol{k}}^{\mu}$ as well as the dipole transition matrix elements $\mathbf{d}_{\boldsymbol{k}}^{\mu\nu}$ to compute the change in the electronic occupation numbers $n_{\boldsymbol{k}}^{\mu}$ according to
\begin{equation}
\begin{split}
    \partial_{t}n_{\boldsymbol{k}}^{\mu}&
     =\frac{2\pi}{\hbar}\sum_{\nu} |\mathcal{E}(t)|^2 \,  \big|\hat{\mathbf{e}}_{\alpha}\cdot\mathbf{d}_{\boldsymbol{k}}^{\mu\nu}\big|^2 \\
     & \qquad\qquad \times g\left(\epsilon_{\boldsymbol{k}}^{\mu}-\epsilon_{\boldsymbol{k}}^{\nu}\pm\hbar\omega_{\mathrm{L}}\right)(n_{\boldsymbol{k}}^{\nu}-n_{\boldsymbol{k}}^{\mu})
     \label{eq:dn-dt}
\end{split}
\end{equation}
for dipole transitions $\vert \boldsymbol{k},\mu\rangle\to\vert\boldsymbol{k},\nu\rangle$ \cite{essert_electron-phonon_2011,Stiehl2022}. The optical field is characterized by a time dependent amplitude~$\mathcal{E}(t)$ and a polarization vector~$\hat{\mathbf{e}}_{\alpha}$. The broadening of the optical transition energies is modeled by a  Gaussian~$g(\Delta\epsilon)$. Solving Eq.~\eqref{eq:dn-dt} for the pulse parameters given in App.~\ref{ch:dielectric} leads to the excited carrier distributions created by the pulse. We denote the distributions after pulse excitation by~$n_{\boldsymbol{k}}^{\mu}(\hat{\mathbf{e}}_\alpha)$ to explicitly show their dependence on the pump polarization vector $\hat{\mathbf{e}}_{\alpha}$.
 
Figure~\ref{fig:fig2}(a) shows the polarization vector $\hat{\mathbf{e}}_{\alpha}$ of the linearly polarized pulse in a plane perpendicular to the propagation direction. The angle~$\alpha$ determines the direction of the polarization as measured against the azimuthal unit vector~$\hat{\mathbf{e}}_{\theta}$. The rotation around a direction $\hat{\mathbf{n}}$ is given, in general, by a matrix
\begin{equation}
    R(\hat{\mathbf{n}},\alpha) = [1-\cos(\alpha)]\,\hat{\mathbf{n}}\otimes\hat{\mathbf{n}}+\cos(\alpha)\, \mathds{1}+\sin(\alpha)\,\big[\hat{\mathbf{n}}\big]_{\times}.
\end{equation}
Here $\otimes$ indicates a dyadic product, $[\hat{\mathbf{n}}]_{\times}$ a cross product matrix and $\mathds{1}$ the $3\times3$ identity matrix.
For an incident direction parametrized by spherical coordinates $\hat{\mathbf{e}}_{r}=(\sin\theta\cos\varphi,\sin\theta\sin\varphi,\cos\theta)^T$, the corresponding polarization vector is
\begin{equation}
    \hat{\mathbf{e}}_{\alpha}= R(\hat{\mathbf{e}}_r,\alpha)\hat{\mathbf{e}}_{\theta} = \cos(\alpha) \hat{\mathbf{e}}_{\theta} +\sin(\alpha) \hat{\mathbf{e}}_{\varphi}. 
\end{equation}
We will always use Fig.~\ref{fig:fig2}(a) for the definition of angles and vectors in the following. 

To characterize the influence of the optical excitation on the spin degree of freedom, we define the ($y$ component of the) \emph{total} spin polarization 
\begin{equation}
    S_{\mathrm{tot}}^{y}(\hat{\mathbf{e}}_\alpha)=\frac{2\mu_{\mathrm{B}}}{\hbar}\frac{1}{N_k}\sum_{\boldsymbol{k},\mu}n_{\boldsymbol{k}}^{\mu}(\hat{\mathbf{e}}_\alpha)\langle s_{\boldsymbol{k},\mu}^{y}\rangle
\end{equation}
and the \emph{excited} spin polarization 
\begin{equation}
    S_{\mathrm{exc}}^{y}(\hat{\mathbf{e}}_\alpha)=\frac{2\mu_{\mathrm{B}}}{\hbar}\frac{1}{N_{k}}\sum_{\substack{\boldsymbol{k},\mu \\ \epsilon_{\boldsymbol{k}}^{\mu}>E_{\mathrm{F}}}}n_{\boldsymbol{k}}^{\mu}(\hat{\mathbf{e}}_\alpha)\langle s_{\boldsymbol{k},\mu}^{y}\rangle 
    \label{eq:s-y}
\end{equation}
where $n_{\boldsymbol{k}}^{\mu}(\hat{\mathbf{e}}_\alpha)$ are the carrier distributions created by the optical pulse with polarization vector $\hat{\mathbf{e}}_{\alpha}$. Further, the spin expectation values $\langle s_{\boldsymbol{k},\mu}^{y}\rangle$ were introduced in Sec.~\ref{ch:section2}. The excited spin polarization arises exclusively from the pulse-excited population of electronic states above the Fermi energy and is therefore essentially a non-equilibrium quantity. Note that the total spin polarization can only be changed in the presence of SOC-induced spin mixing in the electronic states connected by an optical transition. 

We focus next on two representative directions in the BZ, which are shown in Figs.~\ref{fig:fig1}(d) and (e) to illustrate the distinctive optical responses of a g-wave \emph{bulk} altermagnet. First we investigate the case of normal incidence when the propagation direction of the laser pulse is oriented along the $[001]$-direction of the crystal. In order provide a simple picture of this case, we examine the spin cut of the plane perpendicular to the incident direction, which is divided by the nodal planes into six triangular areas, cf.~Fig.~\ref{fig:fig1}(d). We assume an above-gap excitation with a photon energy of $3.75\;\mathrm{eV}>1.44\;\mathrm{eV}=E_{\mathrm{gap}}$ and an excitation angle of $\alpha=0^{\circ}$ (purple polarization vector). Figure~\ref{fig:fig2}(b) shows the computed pulse-excited electronic distribution in this plane in the first conduction band. 
The color code for the excited carrier occupation numbers ranges from zero in white to 0.5 or more indicated by red, the nodal planes are highlighted by black lines. The electronic distributions are highly anisotropic and opposite areas in a perpendicular plane exhibit identical electron distribution functions, i.e., $n(k_x,k_y,k_z)=n(-k_x,-k_y,k_z)$. 
 
Figure~\ref{fig:fig3}(a) contains the excited spin polarization obtained from Eq.~\eqref{eq:s-y} and the solution of the dynamical equation~\eqref{eq:dn-dt}. The spin polarization is obtained from the integration over the whole BZ and gives a vanishing result for all excitation angles including the special case shown in Figs.~\ref{fig:fig2}(b) and (c). 
From the spin texture and excited electron distribution of Figs.~\ref{fig:fig2}(b) and (c) the vanishing \emph{excited} spin polarization for $\alpha =0$ can be easily visualized. Figure~\ref{fig:fig3}(a) shows further that a null result is obtained for \emph{all} excitation angles, which is in agreement with the spin cut shown for this incident direction of $\theta=0^{\circ}$ in inset i) in Fig.~\ref{fig:fig3}(a): Any excitation angle $\alpha$ should lead to an equal excitation of up and down spins in the plane of the spin cut, and thus to a vanishing spin polarization. This expectation based on the spin-cut  picture is borne out by full calculation, which show the usefulness of spin cuts. The takeaway from the calculation and the spin-cut picture is that it is impossible to excite a spin polarization with a laser pulse impinging in the $[001]$-direction. The absence of an excited spin polarization for all pump polarization angles is a significant difference of $\alpha$-MnTe compared to the planar d-wave altermagnets~\cite{weber_optics_ruo2} RuO$_2$ and KRu$_4$O$_8$.

Next we turn to the case of an incident direction with $\theta=90^{\circ}$ along the $[110]$-axis of the crystal ($x$-direction), as illustrated in Fig.~\ref{fig:fig1}(e). The computed spin polarization obtained from the integration over the whole BZ for this case is also shown in Fig.~\ref{fig:fig3}(a), where, for the same photon energy as before, the excitation angle $\alpha$ is varied between 0$^{\circ}$ and 180$^{\circ}$. The signal shows an angle dependence that is close to sinusoidal, in particular with zeroes at 0° and 90° and minima/maxima around 60$^{\circ}$ and 120$^{\circ}$, respectively. In order to explain this behavior, we plot the excited carrier distribution in the first conduction band for $\alpha=0^{\circ}$ in Fig.~\ref{fig:fig3}(b) and for $\alpha=45^{\circ}$ in Fig.~\ref{fig:fig3}(c). In both cases, electrons are predominantly excited perpendicular to the polarization vector of the laser pulse, which results again in an anisotropic carrier distribution in the band. For $\alpha=0^{\circ}$ in Fig.~\ref{fig:fig3}(b) this does not lead to an excited spin polarization due to equally populated spin-up and spin-down channels. Relating Fig.~\ref{fig:fig3}(c) for $\alpha=45^{\circ}$ with the corresponding spin cut in Fig.~\ref{fig:fig1}(e) suggests, however, that in this case a non-vanishing excited spin polarization signal should result, which is indeed the result obtained from the summation over the whole BZ. This comparison shows that, again, the overall behavior of the spin signal can be qualitatively well described by the spin cut shown in inset ii) of Fig.~\ref{fig:fig3}(a). 

\begin{figure*}[t]
    \centering
    \includegraphics[width=\linewidth]{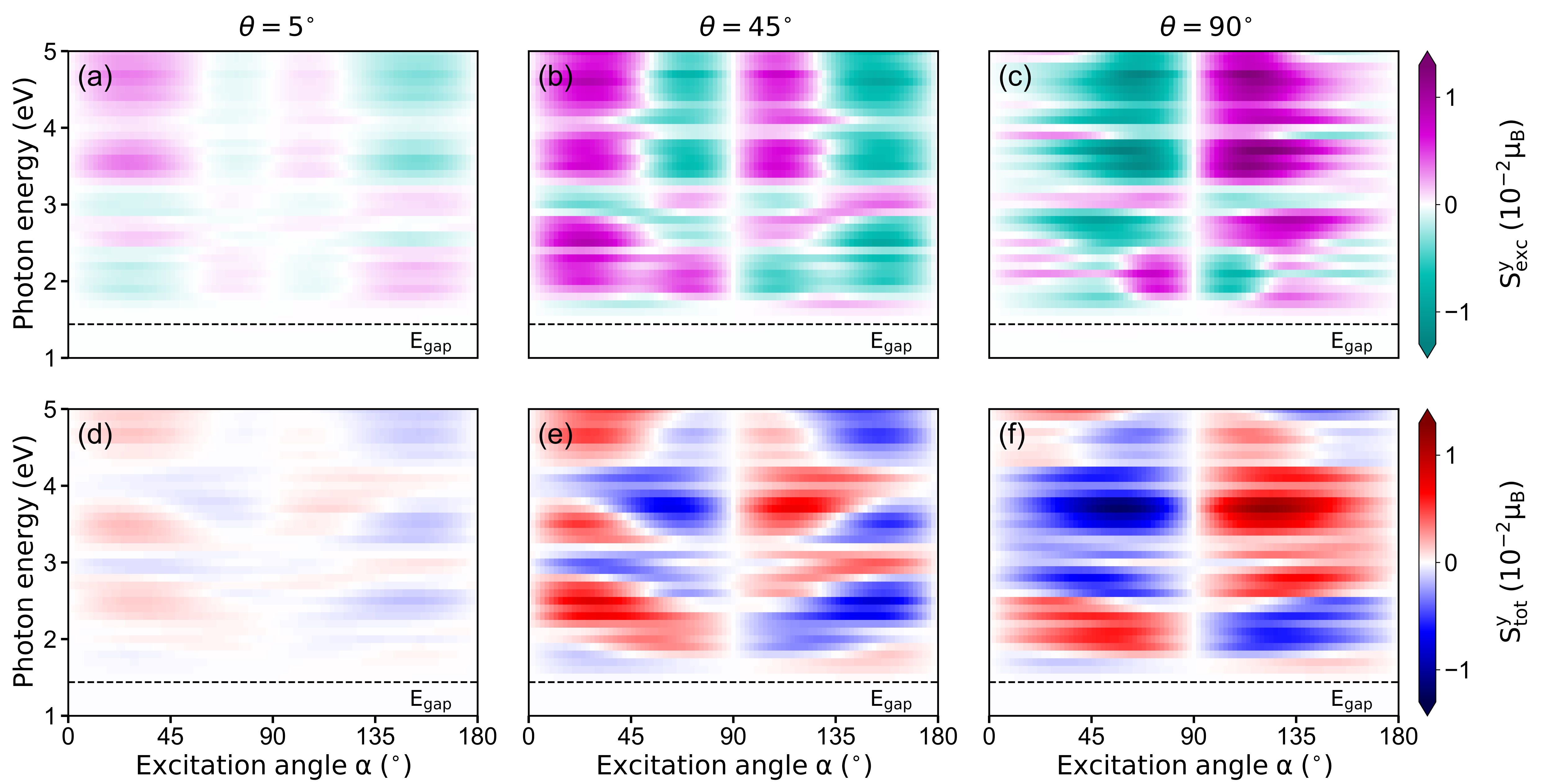}
    \caption{(a-c) Excited spin polarization $\mathrm{S_{exc}^y}$ and (d-f) total spin polarization $\mathrm{S_{tot}^{y}}$ in dependence of the photon energy and the excitation angle for different incident laser directions ranging from an incidence angle of $\theta=5^{\circ}$ (a), (d) over $\theta=45^{\circ}$ (b), (e) to $\theta=90^{\circ}$ (c), (f).}
    \label{fig:fig4}
\end{figure*}

The simple 2D picture introduced above for the spin response also works for other incident directions. For $\theta=48^{\circ}$, Fig.~\ref{fig:fig3}(a) shows a pronounced $\sin(4\alpha)$-dependence of $S_{\mathrm{exc}}^{y}$ with maxima and minima around odd multiples of $22.5^{\circ}$. The corresponding spin cut in inset iii)  provides a visualization of the spin structure of the uppermost valence band in this plane of the BZ perpendicular to the direction of $\theta=48^{\circ}$. The eight alternating spin sectors of the spin cut corresponding to this incident direction are in qualitative agreement with the sign changes of the computed signal. Interestingly, this particular spin cut of the bulk g-wave altermagnet $\alpha$-MnTe resembles the spin cut of a planar g-wave altermagnet.

Figure~\ref{fig:fig3}(d) combines the dependence of the excited spin polarization on incident angles $\theta$ and excitation angles $\alpha$ in one plot. The colormap shows a white, i.e., non-spin polarized, region for $\theta=0^{\circ}$ which corresponds to a vanishing signal for $[001]$ incidence. With increasing incident angle, a spin polarization signal with an oscillating behavior starts to appear for increasing $\theta$. For small $\theta$ up to a transition around 65$^{\circ}$ the signal shows four zero crossings and then changes to two zeros for $\theta$ close to 90$^{\circ}$. Fig.~\ref{fig:fig3}(a) suggests that the signal can essentially be decomposed into a combination of $\sin(2\alpha)$ and $\sin(4\alpha)$ contributions. For comparison, we present the corresponding result without SOC contributions in Fig.~\ref{fig:fig2_sup} in App.~\ref{ch:appendix2}. In this case the excited spin polarization is exactly determined by the nodal-plane alignment of the altermagnet.

In order to make this observation more quantitative, we disentangle the planar d-wave like and planar g-wave like regions by introducing a fit function
\begin{equation}
    f_{\theta}(\alpha)=A^{\theta}_{\mathrm{d}}\sin(2\alpha-\phi^{\theta}_{\mathrm{d}})+A^{\theta}_{\mathrm{g}}\sin(4\alpha-\phi^{\theta}_{\mathrm{g}}) \label{eq:fit_function}
\end{equation}
to the excited spin polarization with amplitudes $A^{\theta}_{\mathrm{d,g}}$ and phases $\phi^{\theta}_{\mathrm{d,g}}$. The fitting procedure and the quality of the fits are presented in detail in the Supplementary Information~S2~\cite{supplement}. 
The resulting amplitudes $A^{\theta}_{\mathrm{d,g}}$ are shown in Fig.~\ref{fig:fig3}(e). For normal incidence both of the fit amplitudes are zero because the signal vanishes. As soon as the incident angle is increased, a signal with d-wave and g-wave like contributions starts to appear, where both amplitudes increase. At around $25^{\circ}$ incidence the d-wave like amplitude starts to decrease again until it is almost completely gone for an incident angle of $48^{\circ}$. Increasing the incident angle further, the d-wave like contribution rises again, while the g-wave like contribution shrinks. Finally, the d-wave like part becomes larger than the g-wave like contribution at approximately $65^{\circ}$. Note that the planar g-wave like contribution is always present in the signal and never vanishes except for $\theta=0^{\circ}$. Such a procedure might also be used in the analysis of experimental results. 

We stress that our calculated ``signals''  in form of an excited/total spin polarization are very likely experimentally accessible in magneto-optical detection schemes in a similar fashion to experiments on RuO$_2$ \cite{weber_optics_ruo2}, even though we have not modeled a complete pump-probe setup. For $\alpha$-MnTe, high-quality patterned single-domain samples exist \cite{Amin2024_nature} with a Néel vector direction in the plane of the sample. In in order to experimentally detect changes of the spin polarization, one therefore would have to choose a longitudinal or transverse MOKE geometry  different from the normal-incidence polar MOKE configuration in Ref.~\cite{weber_optics_ruo2}. Finally, we note that the results discussed here also quite generally apply to changes in the polarization characteristics with respect to spin projection on the Néel vector direction of other bulk g-wave systems such as CrSb. This material  exhibits [001] Néel order~\cite{Reimers2024}, which should allow for a polar probe configuration.

\section{Optical Excitation Conditions: Photon-energy Dependence \label{ch:section4}}

We now address the dependence of the optically excited spin polarization on the laser photon energy or, equivalently, the wavelength. We intend to show that the angle dependence discussed so far is rather general for wavelengths in the visible spectral range. We wish to analyze, in particular, the cases with slightly off-normal incidence ($\theta=5^{\circ}$), an intermediate angle of incidence with $\theta=45^{\circ}$, where we expect a dominant planar g-wave like dependence and the case with $\theta=90^{\circ}$, where we expect a dominant planar d-wave like behavior. We account for a change in absorption of the laser pulse due to the wavelength dependent refractive index as described in App.~\ref{ch:dielectric}. Figure~\ref{fig:fig4} summarizes the results for laser photon energies $\hbar\omega_{L}=1$\,eV to 5\,eV. The upper row, Fig.~\ref{fig:fig4}(a)-(c), shows the excited spin polarization $S_{\text{exc}}^{y}$ with a color scheme ranging from negative (turquoise) to  positive (purple). The lower row, Fig.~\ref{fig:fig4}(d)-(f), shows the total spin polarization $S_{\text{tot}}^{y}$ indicated by colors ranging from blue (negative spin polarization) to red (positive spin polarization). In addition we indicate the energy of the band gap ($E_{\mathrm{gap}}$) of $\alpha$-MnTe by a black dashed line to show that no spin polarization is excited at all for an excitation with an energy below the band gap. 
For an incident angle close to normal incidence, i.e., $\theta=5^{\circ}$, Figs.~\ref{fig:fig4}(a),(d) show only a small signal in the excited and total spin polarization for all photon energies. In Figs.~\ref{fig:fig4}(b),(e) and (c),(f)  for the cases of $\theta=45^{\circ}$ and $\theta=90^{\circ}$, one can identify the $\sin(4\alpha)$ and $\sin(2\alpha)$ behavior, respectively, as discussed in Sec.~\ref{ch:section3}. This expected behavior is found for a wide range of photon energies, but discrepancies from the simple picture based on the spin cuts, as shown in the insets in Fig.~\ref{fig:fig3}(a), may occur due to the orbital character of the states involved in the optical transitions and the number of excited bands. This is illustrated by the photon energy range from $E_{\mathrm{gap}}$ up to 2.4\,eV, where the spin polarization is determined exclusively by the two topmost valence bands and the first two conduction bands,  which show an opposite spin character, cf.~Fig.~\ref{fig:fig1}(c) and Supplementary Information S1~\cite{supplement}. For such an excitation the spin polarization is very sensitive to the individual contributions from each of the few bands involved. For higher photon energies, more bands are excited and the spin polarization is more likely to be determined by the general spin structure of the bulk g-wave altermagnet (see Supplementary Information S1~\cite{supplement}), which is reflected in the spin cuts. Note that a sign change occurs in the spin polarization around 3\,eV since bands with opposite spin character are excited. This sign change is an intrinsic altermagnetic feature as it can also be observed in the excited spin polarization calculated without SOC, which is shown in Fig.~\ref{fig:fig1_sup}(b) and (c) in App.~\ref{ch:appendix2}. The total spin polarization in Fig.~\ref{fig:fig4}(d)--(f) shows qualitatively similar features to those discussed for the excited spin polarization, but these are absent in Fig.~\ref{fig:fig1_sup}(e) and (f) computed without SOC. This illustrates that it is possible to create a transient magnetization in the sample with optical pulses in the presence of SOC as determined by the altermagnetic symmetries.  

\begin{figure*}[t]
    \centering
    \includegraphics[width=\linewidth]{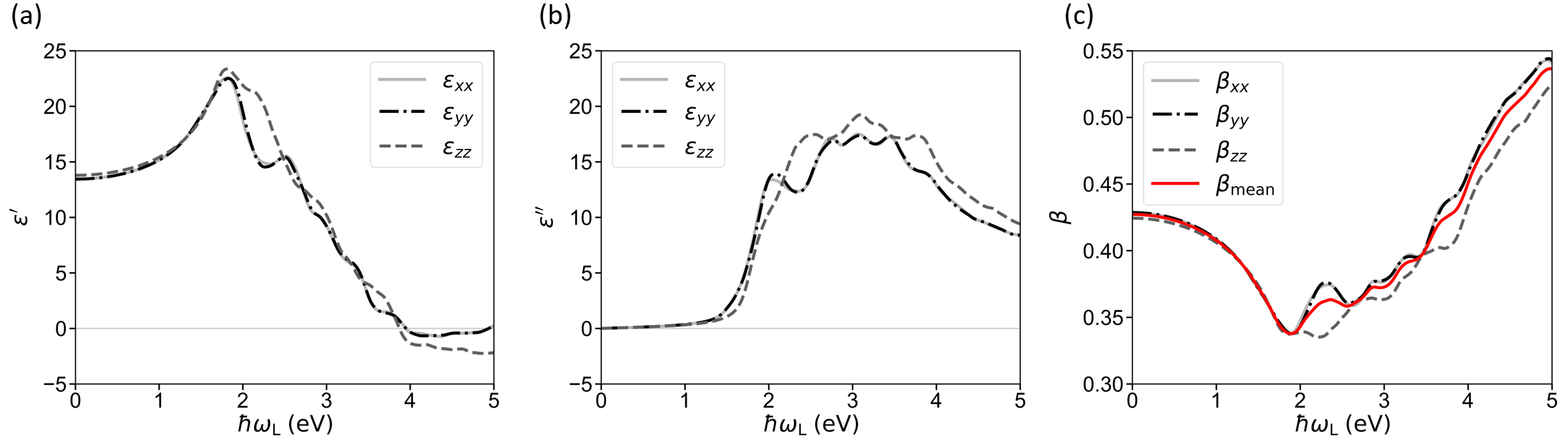}
    \caption{(a) Real and (b) imaginary part of the diagonal dielectric tensor elements for bulk $\alpha$-MnTe. (c) corresponding reduction factor for the electric field amplitude.}
    \label{fig:dielectric_function}
\end{figure*}

\section{Conclusion \label{ch:section5}}

We studied the characteristics of  electronic spin-polarizations driven by linearly polarized optical pulses in the prototypical bulk g-wave altermagnet $\alpha$-MnTe. The spin dependence of Bloch states of this altermagnetic class together with the anisotropic excited electron distributions in $k$-space make a purely optical control of spins in such a magnetically compensated material possible.
In difference to planar d-wave systems, the angle of incidence of the excitation pulse has an important influence in bulk g-wave altermagnets. Depending on this angle, the induced spin polarization shows different dependences on the direction of the linear polarization of the exciting laser pulse, but the six-fold symmetry of the crystal structure is never reflected in the excited spin polarization. For special angles of incidence the periodicity of the induced spin polarization even resembles that of a planar d-wave system. The induced electronic spin polarizations were obtained from calculations involving electronic excitations in the whole BZ and microscopic quantities such as the dipole moments. Nevertheless the dependence of the spin polarization on the excitation conditions can qualitatively be well predicted in a simple and straightforward way from ``spin cuts'', i.e., the dependence of the  spin expectation values of Bloch states in a plane perpendicular to the incident direction of the laser pulse. In addition, we showed that experiments that measure the telltale dependence of the non-equilibrium electronic spin polarization on the optical polarization can serve as a probe of the bulk altermagnetic character. Our results therefore may motivate and guide further time-dependent magneto-optical experiments on altermagnetic candidate materials. 

\begin{acknowledgments}
Funded by the Deutsche Forschungsgemeinschaft (DFG, German Research Foundation) – TRR 173 – 268565370 (projects A03 and A08).
\end{acknowledgments}

\section*{Data Availability}
The data and computer codes are not publicly available. They are available from the authors upon reasonable request.

\appendix

\section{Dielectric Function \label{ch:dielectric}}

To characterize the wavelength dependence of the absorption and reflection in $\alpha$-MnTe, we calculated the dielectric tensor $\epsilon_{\alpha\beta}(q\to 0,\omega)$ within the random phase approximation (RPA) in the wavelength range from $\hbar\omega_{\mathrm{L}}=0$\,eV to $\hbar\omega_{\mathrm{L}}=5$\,eV. To eliminate artifacts resulting from the discrete mesh,  $32\times 32\times 20$ were used together with a smearing of 0.003\,Ha. Figures~\ref{fig:dielectric_function}(a),(b) show the real and imaginary parts of different dielectric tensor components, respectively. These results qualitatively match those of Refs.~\cite{mazin_2023_dielectric_function,oleszkiewicz19881}, showing a peak in the real part of the dielectric function around 2\,eV. The birefringent character of hexagonal $\alpha$-MnTe is apparent as two of the diagonal components are almost completely identical, $\epsilon_{xx}=\epsilon_{yy}\neq\epsilon_{zz}$. From the dielectric function we calculate the complex refractive index $\tilde{n}(\omega)=n(\omega)+i\kappa(\omega)$, where $n(\omega)$ is the real refractive index and $\kappa(\omega)$ the extinction coefficient. Our main goal here is to take into account the effect of the reduction of the electric field amplitude in the material in  in order to have a meaningful comparison with the field strength/intensity of the pump pulse in experiments. We do not attempt a complete model of an experimental pump-probe setup and thus assume, for simplicity, an effective refractive index that results from the mean value of the three Cartesian directions $\bar{n}=(n_{xx}+n_{yy}+n_{zz})/3$. Based on this assumption, the factor
\begin{equation}
    \beta_{\mathrm{mean}}(\omega)=\frac{2}{\sqrt{[1+\bar{n}(\omega)]^2+\bar{\kappa}(\omega)^2}}
\end{equation}
determines the reduction of the electric field amplitude $E_{0}^{\prime}(\omega)=E_{0}\beta(\omega)$ in the material at a given wavelength. It is plotted in Fig.~\ref{fig:dielectric_function}(c). 

The laser pulse is modeled by a Gaussian envelope
\begin{equation}
    \begin{split}
    \mathbf{E}(t) & =  E^{\prime}_{0}(\omega_\mathrm{L})\exp\left[-4\ln(2)\frac{t^2}{\tau_{\mathrm{FWHM}}^2}\right]\hat{\mathbf{e}}_{\alpha}g(\Delta \epsilon)\\
     & = \mathcal{E}(t) \hat{\mathbf{e}}_{\alpha} g(\Delta \epsilon)
    \end{split}
\end{equation}
where $\omega_{\mathrm{L}}$ is the laser frequency, $\hat{\mathbf{e}}_{\alpha}$ the polarization vector defined in Sec.~\ref{ch:section3} and $\tau_{\mathrm{FWHM}}=40$\,fs the full width at half maximum. The energetic broadening of $\Gamma=100$\,meV is introduced by
\begin{equation}
	g(\Delta\epsilon)=\frac{1}{\sqrt{2\pi}}\frac{\sqrt{4\ln(2)}}{\Gamma}\exp\left[-4\ln(2)\frac{\left(\vert\epsilon_{\boldsymbol{k}}^{\mu}-\epsilon_{\boldsymbol{k}}^{\nu}\vert-\hbar\omega_{\mathrm{L}}\right)^2}{\Gamma^2}\right].
\end{equation}
The pump fluence is determined by
\begin{equation}
    \rho_{\mathrm{E}}=\frac{1}{2}\epsilon_{0}c\int_{-\infty}^{\infty}\vert \mathbf{E}(t)\vert^2\mathrm{d}t=\frac{1}{4}E_{0}^2\epsilon_{0}c\sqrt{\frac{\pi}{2\ln(2)}}\tau_{\mathrm{FWHM}}\label{eq:fluence}.
\end{equation}
Thus we calculate the electric field amplitude for a fixed fluence of $\rho_{E}=4\;\frac{\mathrm{mJ}}{\mathrm{cm}^2}$ according to
\begin{equation}
    E_{0}=\left[\frac{4\rho_{\mathrm{E}}}{\epsilon_{0}\tau_{\mathrm{FWHM}}c}\sqrt{\frac{2\ln(2)}{\pi}}\right]^{1/2}.
\end{equation}

\section{Excitation without SOC \label{ch:appendix2}}
\begin{figure}
    \centering
    \includegraphics[width=\linewidth]{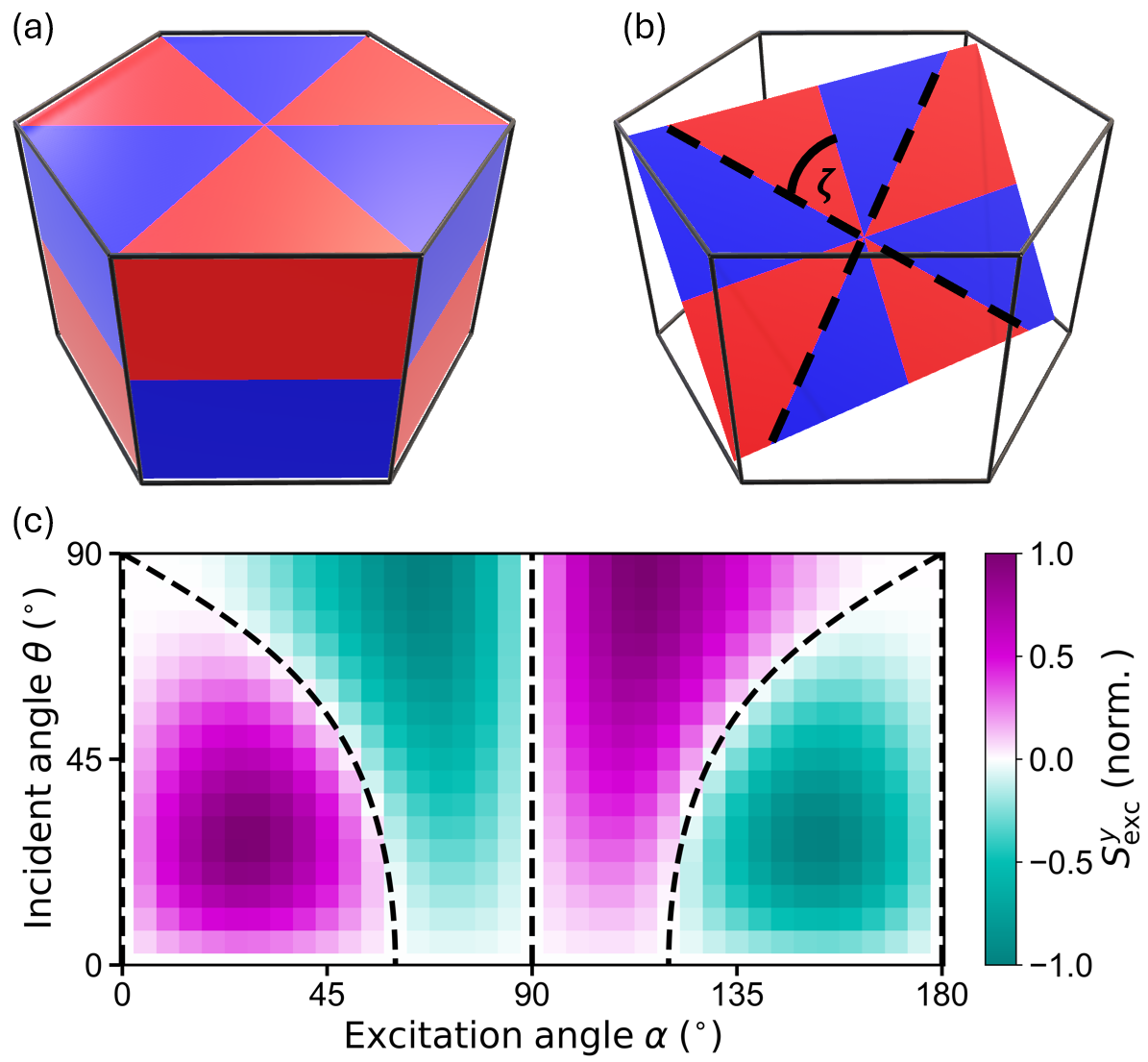}
    \caption{(a) Schematic spin arrangement of the bulk g-wave altermagnet in the entire 3D BZ, (b) spin cut through the BZ, with the nodal plane angle~$\zeta$ highlighted. (c) Computed excited spin polarization without SOC for a photon energy of $\hbar\omega_{\mathrm{L}}=3.50$~eV for different incident and excitation angles. Dashed black lines represent the excitation angles at which a nodal plane is observed in the 2D cut of the BZ, perfectly matching the roots in the spin polarization signal.}
    \label{fig:fig2_sup}
\end{figure}
\begin{figure*}
    \centering
    \includegraphics[width=\linewidth]{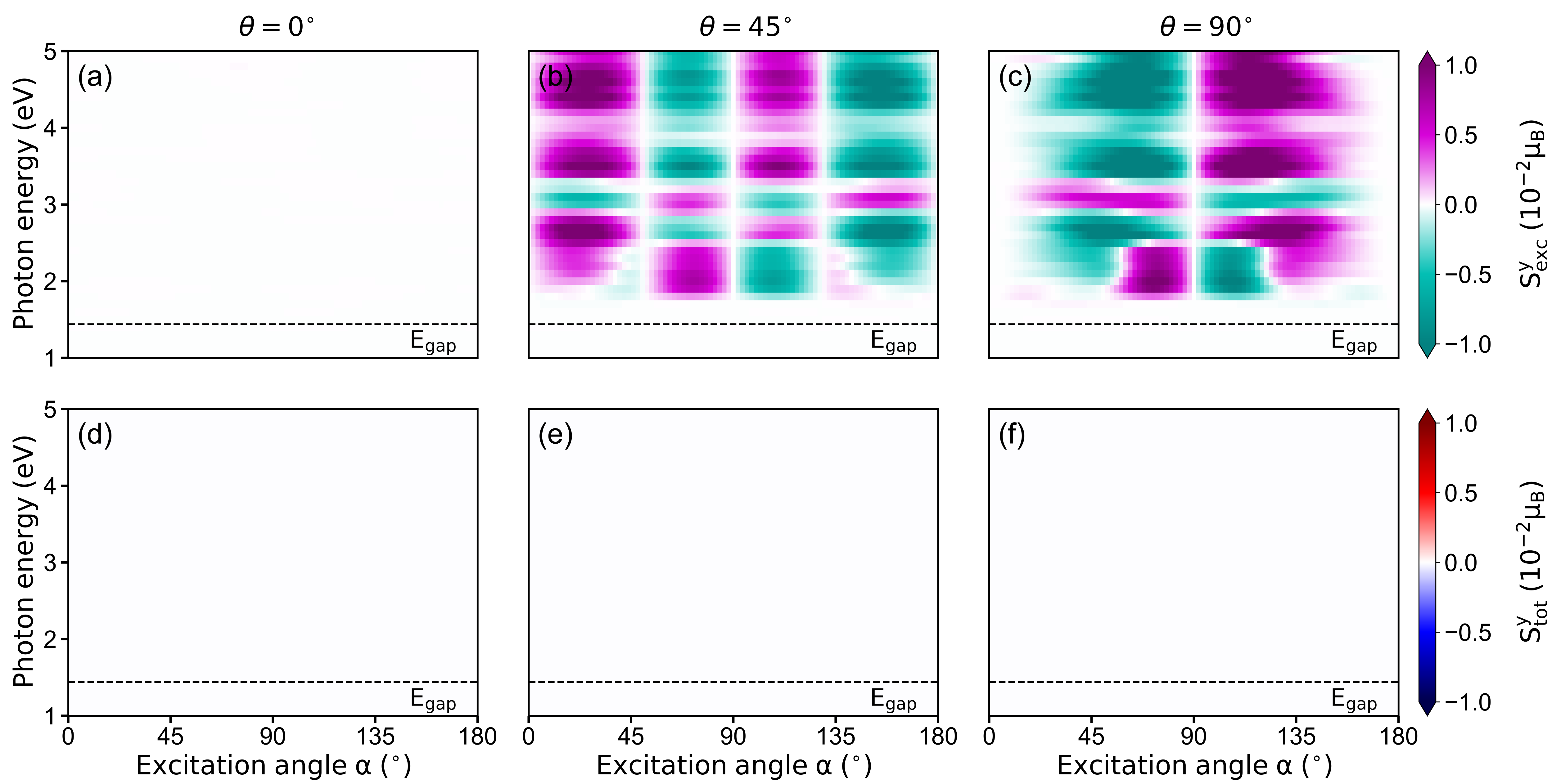}
    \caption{Spin polarization as a function of photon energy computed without SOC. (a)--(c) Excited spin polarization and (d)--(f) total spin polarization for different incident angles. The white color in subfigures (a) and (d)--(f) highlights a vanishing spin polarization signal.}
    \label{fig:fig1_sup}
\end{figure*}
In this appendix we present results computed without SOC in order to show that the spin polarization induced in the excited states and its symmetries do not arise due to the influence of SOC. We first focus on the position of the roots in the spin polarization signal. Referring to the schematic spin structure in the bulk g-wave altermagnet in Fig.~\ref{fig:fig2_sup}(a), we can track the position of the nodal lines, i.e., the projection of the nodal planes in the spin cut, as function of the incident angle~$\theta$. Independent of the incident angle, Fig.~\ref{fig:fig2_sup}(b) shows that we always have a nodal line at the excitation angles~$\alpha=0^{\circ}$, $\alpha=90^{\circ}$ and~$\alpha=180^{\circ}$. The additional nodal lines in the 2D cut of the BZ, shown as black dashed line in Fig.~\ref{fig:fig2_sup}(b), changes its position depending on~$\theta$. 
In the case without SOC the excitation angle~$\zeta$ at which this nodal line occurs can be analytically determined as
\begin{equation}
\zeta(\theta)=\arccos\left(\frac{1}{\sqrt{1+3\cos^2(\theta)}}\right).
\end{equation}
In Fig.~\ref{fig:fig2_sup}(c) we calculated the dependence of the excited spin polarization on the incident angle for a fixed photon energy of $\hbar\omega_{\mathrm{L}}=3.50$ eV without SOC contributions according to~\eqref{eq:s-y}. The color code used here is the same as for the excited spin polarization in the main manuscript. The black dashed lines show the positions of the nodal lines at $\alpha=0^{\circ}$, $\alpha=90^{\circ}$, $\alpha=180^{\circ}$, $\alpha=\zeta(\theta)$ and $\alpha=180^{\circ}-\zeta(\theta)$. One can see that the nodal line positions perfectly match the zeros of the excited spin polarization, thus highlighting the remarkable agreement between the excited spin polarization and the bulk g-wave spin order. 

Figure~\ref{fig:fig1_sup} shows the photon-energy dependence of the (a)--(c) excited and (d)--(f) total spin polarization obtained from an \emph{ab initio} calculation without SOC, which should be compared with Fig.~\ref{fig:fig4} in the main manuscript. In difference to Fig.~\ref{fig:fig4} in the main text, where SOC is included, excitations are now only possible by a spin conserving transition which is reflected in the vanishing total spin polarization as indicated by the white color in Fig.~\ref{fig:fig1_sup}(d)--(f).
In Fig.~\ref{fig:fig1_sup}(a), for normal incidence, the excited spin polarization vanishes for every wavelength and polarization angle, as expected by the symmetries discussed in the main text. In Fig.~\ref{fig:fig1_sup}(b) we changed the incidence angle to $45^{\circ}$, which leads to  a spin polarization with the expected $\sin(4\alpha)$-like behavior for most of the photon energies. Compared to the case with SOC, the main difference is that the angles for which the spin polarization vanishes do not change for all photon energies considered here. As in the case with SOC, there are discrepancies in the excited spin polarization from the expected symmetries in the energy range from $E_{\mathrm{gap}}$ to about 2.5\,eV due to the small number of bands involved in the corresponding transitions.

\newpage

%

\end{document}


\title{Supplement to \\
Signatures of bulk g-wave altermagnetism in optically excited $\alpha$-MnTe}

\author{Luca~Felipe~Haag}
        \affiliation{Department of Physics and State Research Center OPTIMAS, RPTU University Kaiserslautern-Landau, 67663 Kaiserslautern, Germany}
\author{Marius~Weber}
        \affiliation{Institute of Physics, Johannes Gutenberg University Mainz, 55099 Mainz, Germany}
        \affiliation{Department of Physics and State Research Center OPTIMAS, RPTU University Kaiserslautern-Landau, 67663 Kaiserslautern, Germany}
\author{Kai~Leckron}
    \affiliation{Department of Physics and State Research Center OPTIMAS, RPTU University Kaiserslautern-Landau, 67663 Kaiserslautern, Germany}
\author{Libor~\v{S}mejkal}
    \affiliation{Max Planck Institute for the Physics of Complex Systems, 01187 Dresden, Germany}
    \affiliation{Max Planck Institute for Chemical Physics of Solids,  01187 Dresden, Germany}
        \affiliation{Institute of Physics, Academy of Sciences of the Czech Republic, Cukrovarnick\'{a} 10, Prague 6, Czech Republic}
\author{Jairo~Sinova}
    \affiliation{Institute of Physics, Johannes Gutenberg University Mainz, 55099 Mainz, Germany}
        \affiliation{Department of Physics, Texas A\&M University, College Station, Texas 77843, USA}
\author{Hans~Christian~Schneider}
    \affiliation{Department of Physics and State Research Center OPTIMAS, RPTU University Kaiserslautern-Landau, 67663 Kaiserslautern, Germany}
    
\date{\today}
\maketitle
\newpage

\section{Excitation along a high symmetry path \label{ch:supplement1}}

\begin{figure*}[h!]
    \centering
    \includegraphics[width=\linewidth]{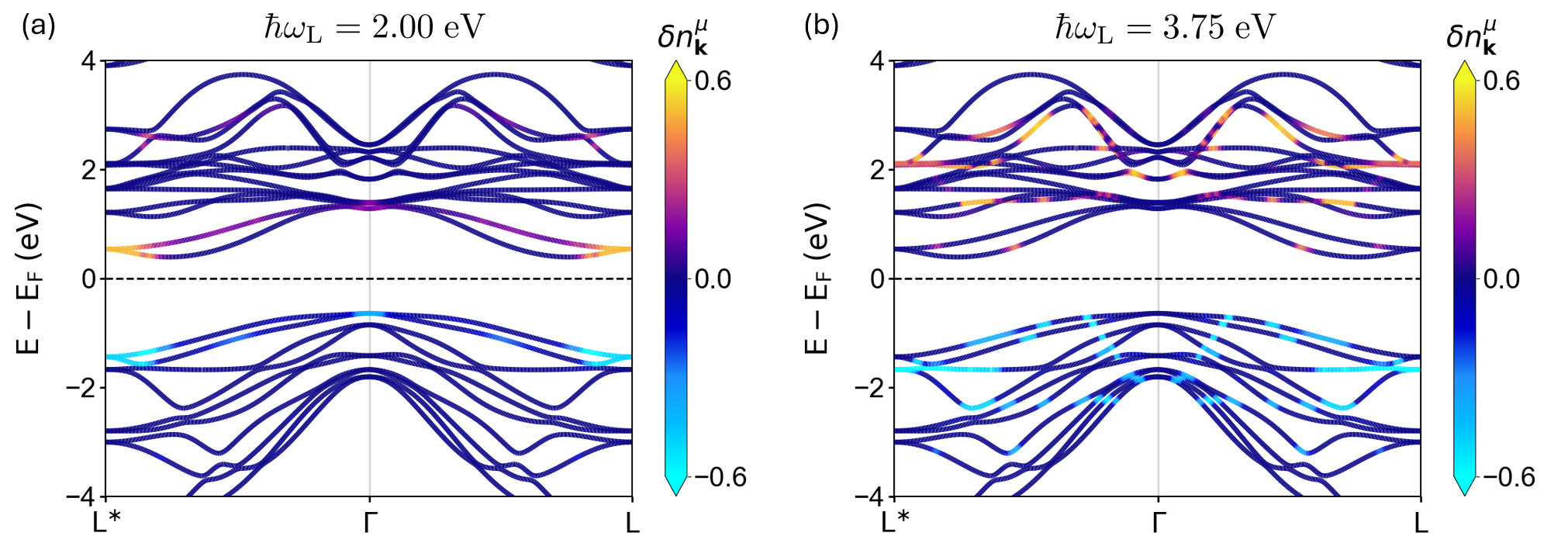}
    \caption{Change in the electronic occupation when exciting $\alpha$-MnTe with a laser pulse with normal incidence and $\alpha=0^{\circ}$. Shown is the change along the 1D-path $L^{\star}-\Gamma-L$ for a photon energy of (a) $\hbar\omega_{\mathrm{L}}=2.00$ eV and (b) $\hbar\omega_{\mathrm{L}}=3.75$ eV to illustrate the number of excited bands. Here, SOC is included.}
    \label{fig:fig1_sup}
\end{figure*}

In the main text we demonstrated that the excited spin polarization differs from the expected behavior for photon energies in the range of 1.44\,eV to roughly 2.5\,eV. To check why these deviations appear we take a closer look at the optical transitions that occur for two different photon energies along high symmetry path $L^{\star}-\Gamma-L$. Figure~\ref{fig:fig1_sup} shows the change in the electronic occupation~$\delta n_{\mathbf{k}}^{\mu}$ with the loss of electronic distribution in the valence bands indicated by cyan and the increase of electronic distribution in the conduction band indicated by yellow.
Fig.~\ref{fig:fig1_sup}~(a) shows that for a photon energy of $\hbar\omega_{\mathrm{L}}=2.00$ eV the main contributions involve only transitions into the first two conduction bands. When computing the excited spin polarization signal, it is largely affected by the individual contributions from these two bands. The optical transition matrix elements take into account the spin and orbital character of the electronic Bloch wave functions computed from DFT so that the dependence on the polarization vector of the electric field in general is non-trivial. In this case the spin polarization signal does not show a direct correspondence to the spin structure of the bulk g-wave altermagnet. Figure~\ref{fig:fig1_sup}~(b) demonstrates that the excitation with a larger photon energy, e.g. $\hbar\omega_{\mathrm{L}}=3.75$ eV, involves significant contributions from numerous individual bands to the excited spin polarization, leading to a signal that corresponds well to the underlying bulk g-wave spin structure.

\section{Sinusoidal Fits \label{ch:supplement2}}

\begin{figure*}[p]
    \centering
    \includegraphics[width=\linewidth]{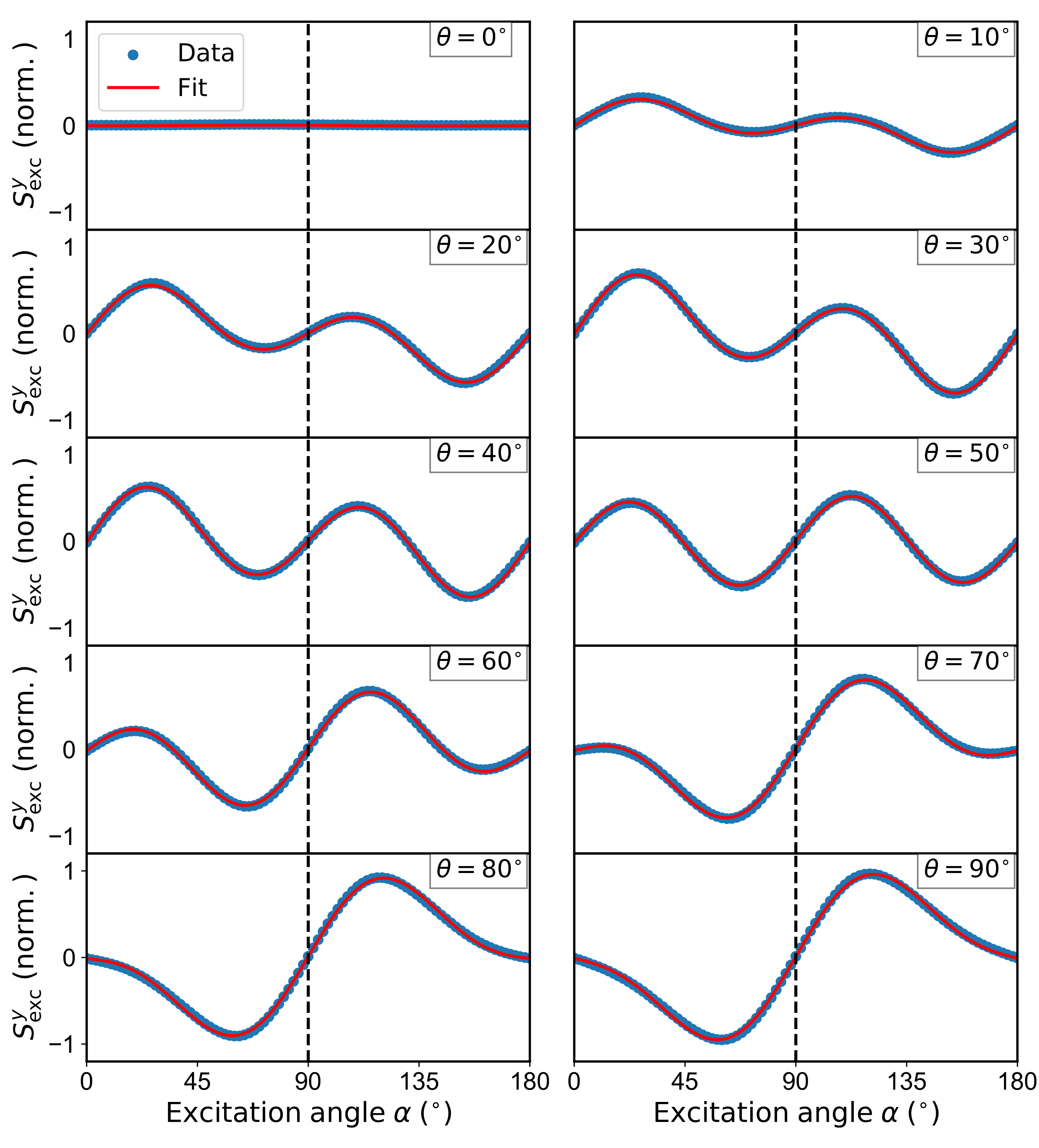}
    \caption{Excited spin polarization for different incident angles $\theta$. The data computed for a photon energy of $\hbar\omega_{\mathrm{L}}=3.75$ eV is shown as blue dots. The sinusoidal fit functions $f_{\theta}(\alpha)$ are shown in red.}
    \label{fig:fig2_supplement}
\end{figure*}

In this section we discuss the sinusoidal fit functions
\begin{equation*}
    f_{\theta}(\alpha)=A^{\theta}_{\mathrm{d}}\sin(2\alpha-\phi^{\theta}_{\mathrm{d}})+A^{\theta}_{\mathrm{g}}\sin(4\alpha-\phi^{\theta}_{\mathrm{g}})   
\end{equation*}
that have been applied to the data computed in the main manuscript in more detail. Figure~\ref{fig:fig2_supplement} illustrates the excited spin polarization for a photon energy of $\hbar\omega_{\mathrm{L}}=3.75$\,eV calculated using our optical model (blue dots), for various incident angles of the laser pulse. On top of these, the sinusoidal fits are shown as red solid lines. Overall, the fit functions are in very good agreement with the computed data. This indicates that the main features of the spin polarization induced in bulk g-wave altermagnets can be captured well by this simple model that takes into account planar d-wave like and planar g-wave like signatures, i.e., typical angular dependencies of the excited spin polarization that are obtained for planar d-wave or g-wave altermagnetic crystal structures. 

\section{Spin-orbit coupling induced lifting of degeneracies \label{ch:supplement3}}

\begin{figure*}[p]
    \centering
    \includegraphics[width=\linewidth]{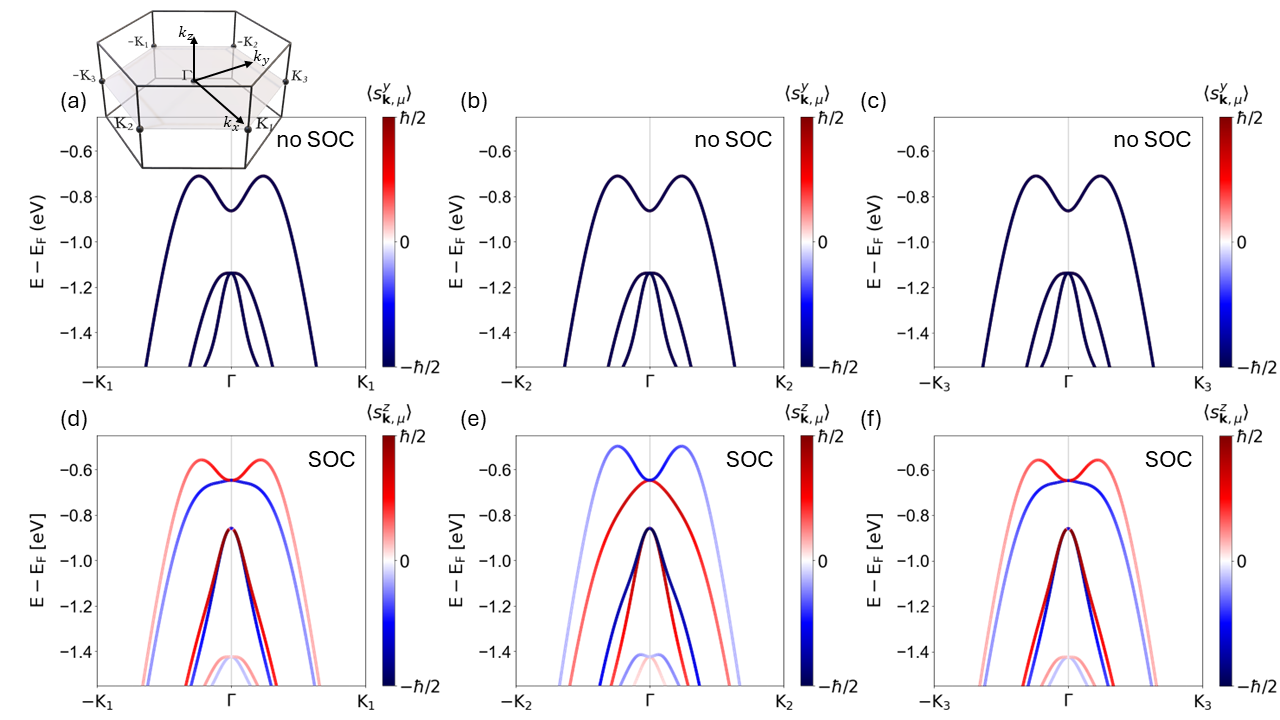}
    \caption{Band structure of $\alpha$-MnTe along high-symmetry paths that connect the various K-points of the Brillouin zone. (a)-(c) shows the calculation neglecting spin-orbit coupling while (d)-(f) shows the bands with SOC highlighting the lifting of the spin degeneracy. Note that in the upper row of the plot the spin-pojection onto the Néel vector directions ($y$-direction) is shown and in the lower row the spin-projection along the $z$-direction is plotted as a color code.}
    \label{fig:fig3_supplement}
\end{figure*}

Here we give a quantitative discussion of the the impact of spin-orbit coupling, which results in a lifting of the spin degeneracy along nodal planes. For this purpose we focus on the plane located at $k_z=0$ and plot the 1D band structure along high-symmetry paths connecting the K-points of the BZ, which can be compared with the Supplementary Information of Ref.~\cite{krempasky2024}. Figure~\ref{fig:fig3_supplement} shows the band structure with color-coded spin expectation value of each Bloch state. In Figs.~\ref{fig:fig3_supplement}(a)-(c) the calculations without SOC are presented where spin-up and spin-down states, defined with respect to the projection along the Néel vector $\langle s_{\boldsymbol{k},\mu}^{y}\rangle$, are degenerate. In this case, the description in terms of nodal planes rigorously applies. In order to show the influence of SOC, we follow Ref.~\cite{krempasky2024} and plot in Figs.~\ref{fig:fig3_supplement}(d)-(f) the spin expectation value along the $z$-direction $\langle s_{\boldsymbol{k},\mu}^{z}\rangle$ as a color code on top of the band structure. The occurrence of this $z$ spin component together with band shifts of several hundred meV demonstrate that the SOC-induced changes of the band structure are significant, but only in limited areas of the BZ. We have already shown in the main part of the manuscript, that the angular dependences of the excited spin polarization signals $S_{\mathrm{exc}}^{y}$ computed from the states in the whole BZ including SOC are described quite well by the nodal plane positions that result from the spin-group description without SOC.

%

